\documentstyle[prd,eqsecnum,aps]{revtex}

\def\bml{\begin{mathletters}}
\def\eml{\end{mathletters}}

\begin{document}

%\preprint{UMP-97/35\hspace{6pt} BGU-97/**\hspace{6pt} RCHEP-97/06}
\draft

%%%%%%%%%%%%%%%%%%%%%%%%%%%%%%%%%%%%%%%%%%%
\twocolumn[                               %
\hsize\textwidth\columnwidth\hsize\csname %
@twocolumnfalse\endcsname                 %
%%%%%%%%%%%%%%%%%%%%%%%%%%%%%%%%%%%%%%%%%%%

\title{Linking Geometric Mass Hierarchy with\\
Threefold Family Replication}
\author{Aharon Davidson and Tomer Schwartz}
\address{Physics Department,
Ben-Gurion University of the Negev,\\
Beer-Sheva 84105, Israel \\
(davidson@bgumail.bgu.ac.il)}
\author{Raymond R. Volkas}
\address{School of Physics,
Research Centre for High Energy Physics,\\
The University of Melbourne,
Parkville 3052, Australia \\
(r.volkas@physics.unimelb.edu.au)}
\maketitle

\begin{abstract}
A link is established between the observed (approximate) geometric
mass hierarchy of quarks and leptons and the triangular structure
of their tenable flavor representations.
This singles out $SU(3)_{H}$ as the horizontal flavor group,
thereby linking the Fermi mass hierarchy with the threefold
family replication.
These linkages are exploited within a flavor-chiral $SU(3)_{H}$
model, with fermions and Higgs bosons in the $\underline{3}
\oplus \underline{6}^{*}$ representation.
The model is Left-Right symmetric and utilizes the
universal see-saw mechanism with a geometric mass suppression
pattern.
Given certain assumptions, the model produces the following
mass-ratio (rather than square-mass-ratio) mixing angle relations
\begin{math}
 {\displaystyle V_{cb}=-V_{ts}=
 \frac{m_{s}}{m_{b}}+\frac{m_{c}}{m_{t}}},
\end{math}
\begin{math}
 {\displaystyle V_{ub}=
 \left(2\frac{m_{c}}{m_{t}}-\frac{m_{s}}{m_{b}}
 \right)\theta_{C}},
\end{math}
\begin{math}
 {\displaystyle V_{td}=
 \left(2\frac{m_{s}}{m_{b}}-\frac{m_{c}}{m_{t}}
 \right)\theta_{C}},
\end{math}
and fixes the ${\displaystyle \frac{m_{u}}{m_{d}}}$ ratio.
\end{abstract}


%%%%%%%%%%%%%%%%%%%%%%%%%%%%%%%%%%%%%%%%%%%
]                                         %
%%%%%%%%%%%%%%%%%%%%%%%%%%%%%%%%%%%%%%%%%%%

\section{Introduction}

Many ideas have been proposed as to how the pattern of quark
and lepton masses and mixings observed in nature might be better
understood. These ideas include
horizontal symmetry\cite{hs1,hs2}, partial\cite{pgut} and full\cite{fgut}
grand unification, radiative mass generation\cite{rmg}, and the
universal see-saw\cite{uss} mechanism.
The latter is a simple Dirac-type extension of the see-saw
mechanism\cite{ss} invented for Majorana neutrinos.
While all of these ideas are very interesting, it is fair to
say that no uniquely compelling model has yet emerged which
provides both a qualitative and a quantitative resolution of
the fermion mass problem within an elegant theoretical framework.

In this paper, we present a theoretical structure for quark masses
and mixings which provides for a qualitative understanding of the
Fermi mass hierarchy, and which suggests five phenomenologically
successful quantitative mass and mixing angle relations.
Our theory is based on an intricate marriage between horizontal
flavor-chiral $SU(3)_{H}$ symmetry and the universal see-saw
mechanism.
This non-trivial amalgam exploits the ability of the universal
see-saw to explain why most quark masses are much lighter than
the electroweak scale together with the ability of spontaneously
broken $SU(3)_{H}$ to discriminate among families.
It thus utilizes the best features of universal see-saw and
horizontal symmetry while avoiding their well-known inadequacies
when used in isolation.
\textit{The core of the model is a strong linkage between the
observed (approximate) geometric mass hierarchy of quarks and
leptons and the specific horizontal symmetry group $SU(3)_{H}$}.
This symmetry group, in turn, is strongly linked to threefold
family replication.

We first discuss the three important hints that are provided by
the combination of mass hierarchy phenomenology and theoretical
ideas:
(i) the Left-Right symmetric universal see-saw mechanism,
(ii) $p$-fold universal see-saw suppression pointing towards
$SU(2)_{H}$ horizontal symmetry, and
(iii) the geometric mass hierarchy and its intimate relation with
$SU(3)_{H}$ horizontal symmetry.
After discussing these hints in Section II, we present in Section
III the theoretical structure that combines all of these ideas and
resolves several apparent incompatibilities among them.
The theory is then used to extract phenomenologically successful
mass and mixing relations for quarks.
We summarize our results in Section IV and discuss some important
open problems.

\section{Hints Towards a Theory of Flavor}

\noindent
{\textbf{Hint 1: Left-Right Symmetric Universal See-saw Mechanism}}

\medskip
Consider the Left-Right symmetric electroweak gauge group
$G_{LR}=SU(2)_{L}\otimes SU(2)_{R}\otimes U(1)_{B-L}$, and
focus, say, on the quark sector.
In addition to the standard left-handed and right-handed quarks
\bml
\begin{eqnarray}
 q_{L} & \sim  & (\underline{2},\underline{1})_{1/3} ~, \\
 q_{R} & \sim & (\underline{1},\underline{2})_{1/3} ~,
\end{eqnarray}
\eml
introduce the exotic vector-like quarks
\bml
\begin{eqnarray}
 Q_{L} & \sim & (\underline{1},\underline{1})_{4/3,-2/3} ~, \\
 Q_{R} & \sim & (\underline{1},\underline{1})_{4/3,-2/3} ~,
\end{eqnarray}
\eml
with matching electric charges.
The structure of the prototype mass matrix is then
\begin{equation}
 \begin{array}{cc}
   & \begin{array}{cc}
     q_{R}\quad & \quad Q_{R}
  \end{array} \\
  \begin{array}{c}
   q_{L} \\  Q_{L}
  \end{array} &
  \left(
  \begin{array}{l|l}
   (\underline{2},\underline{2})_{0} &
   (\underline{2},\underline{1})_{\pm 1}  \\
   \hline
   (\underline{1},\underline{2})_{\pm 1} &
   (\underline{1},\underline{1})_{0}
  \end{array}
  \right)
 \end{array} ~,
\end{equation}
with the various entries denoting the electroweak assignment
of the corresponding Higgs multiplets and/or bare mass terms.
It acquires the universal see-saw form
\begin{equation}
 \left(
 \begin{array}{cc}
  0 & \ell  \\
  r & M
 \end{array}
 \right)
\end{equation}
provided that (i) the putative $(\underline{2}, \underline{2})_{0}$
Higgs multiplet is absent, and (ii) the symmetry breaking chain
is governed by
\begin{equation}
 \ell \ll r \ll M ~,
\end{equation}
where $\ell=\langle(\underline{1},\underline{2})_{\pm 1}\rangle$,
$r=\langle(\underline{2},\underline{1})_{\pm 1}\rangle$, and $M$
stands for the $(\underline{1},\underline{1})_{0}$ bare mass term.
The eigenmasses are approximately given by
\bml
\begin{eqnarray}
 & \displaystyle{m_{light} \simeq \frac{r}{M}\ell} & ~, \\
 & m_{heavy} \simeq M  & ~.
\end{eqnarray}
\eml
Compared with the electroweak scale $\ell$, the light eigenmass
is suppressed by the universal see-saw hierarchy parameter
${\displaystyle \epsilon\equiv\frac{r}{M}}$.
The corresponding light eigenstate, given approximately by
${\displaystyle q-\frac{r}{M}Q}$, exhibits close to standard
electroweak interactions.
Once heavy vector-like leptons are also introduced into the scheme,
the see-saw suppression becomes \textit{universal}.
This universality comes with a bonus. Namely, the Dirac-mass
see-saw suppression in the charged lepton sector automatically
produces\cite{uss} the Gell-Mann--Yanagida see-saw suppression
in the Majorana-neutrino sector.

Although the naive universal see-saw is unable to account for
the full complexity of the observed fermion mass spectrum, it
is possible to extend the idea to incorporate both unsuppressed
and singly-suppressed eigenmasses
\bml
\begin{eqnarray}
 m_{1} & \sim & \ell ~, \\
 m_{2} & \sim & \ell\epsilon ~.
\end{eqnarray}
\eml
This can be achieved through a toy two-family model\cite{toy}
where Left-Right symmetry is accompanied by a horizontal
$U(1)_{H}$ flavor group.
In this model, the full mass matrix has dimension $1+2+1=4$.
Its building blocks (produced when Cabibbo mixing is switched
off) form the Left-Right symmetric sequence
\begin{equation}
 \left(
 \begin{array}{c|cc|c}
   &  &  & \ell  \\
  \hline
   &  & \ell &   \\
   & r & M &   \\
  \hline
  r &  &  &
 \end{array}
 \right) ~.
 \label{toy}
\end{equation}
Note that the unsuppressed eigenmass $\sim\ell$ has, through
Left-Right symmetry, a medium-heavy partner $\sim r$.
The main problem is that such a toy model, while respecting
Left-Right symmetry, cannot be easily modified to also incorporate
a doubly-suppressed eigenmass $\sim\ell\epsilon^{2}$.

\bigskip\noindent
{\textbf{Hint 2: From $p$-fold Suppression to $SU(2)_{H}$}}

\medskip
Motivated by the highly suppressed up and down quark masses,
we now attempt, as an abstract exercise,
to generalize the universal see-saw matrix to
produce a $p$-fold suppressed eigenvalue $\sim\ell\epsilon^{p}$.
The prescription is the following:
Consider a single left-handed standard fermion $\psi_{L0}$ plus
$p$ left-handed see-saw fermions $\Psi_{Li}$, add to it $(p+1)$
right-handed see-saw fermions $\Psi_{Rj}$, and require the
$(p+1)\times(p+1)$ universal see-saw mass matrix to be of the form
\begin{equation}
 \begin{array}{cc}
   & \begin{array}{cccccc}
     \Psi_{Rp} & ~\cdot & ~~\cdot & \Psi_{R2}
     & \Psi_{R1} & \Psi_{R0}
  \end{array} \\
  \begin{array}{c}
   \psi_{L0} \\ \Psi_{L1} \\
   \Psi_{L2} \\ \cdot \\ \cdot \\ \Psi_{Lp}
  \end{array} &
  \left(
        \begin{array}{cccccc}
            & & & & & ~\ell \\
            & & & & ~\chi_{1} & ~M_{1} \\
            & & & ~\chi_{2} & ~M_{2} & \\
            & & \cdot & \cdot & & \\
            &\cdot & \cdot &  & & \\
            ~\chi_{p} & ~M_{p} & &  &
        \end{array}
        \right)
     \end{array} ~.
 \label{bblock}
\end{equation}
The necessary ingredients are that $\ell$ acquires the standard
electroweak scale, and that the $\chi_i$ and $M_j$ are large
$SU(3)_{C}\otimes SU(2)_{L}\otimes U(1)_{Y}$ invariant VEVs or
bare masses with the hierarchy $\chi_{i}\ll M_{j}$.
The ratios ${\displaystyle \epsilon_{i}=\frac{\chi_{i}}{M_{i}}}$
serve as hierarchy parameters in the effective low energy regime.
The $p$--fold suppression of the lightest eigenvalue
\begin{equation}
 m_{light} \simeq
 \ell\epsilon_{1}\epsilon_{2}\ldots
 \epsilon_{p}\sim\ell\epsilon^{p}
\end{equation}
is due to the $p$--fold \textit{stairway} texture of the above
universal see-saw mass matrix.

The above stairway pattern is about to serve as the building block
of the full Fermi mass matrix.
We will consider the full mass matrix shortly, but for the
moment, we need to find an underlying symmetry structure capable
of supporting eq.(\ref{bblock}).
In the 1-fold suppression case, discussed earlier, the underlying
symmetry was Left-Right symmetry.
In the $p$--fold  suppression case ($p>1$), however, a horizontal
symmetry must be invoked to distinguish between the see-saw fermions
in such a way that the prototype stairway pattern is produced.
For a $U(1)_{H}$ symmetry to support such a pattern, the $(p+1)$
right-handed fermions involved (and separately the $p$ see-saw
left-handed fermions) must have their horizontal charges form an
\textit{arithmetic} series.
The horizontal conservation laws
\bml
\label{horcons}
\begin{eqnarray}
  & L_{1}-R_{0}=\ldots=L_{p}-R_{p-1}= a ~, & \\
  &  L_{1}-R_{1}=\ldots=L_{p}-R_{p}= b ~, &
\end{eqnarray}
\eml
for some $a$ and $b$, result from the non-vanishing Yukawa vertices
$\overline{\Psi_{Li}}M_{i}\Psi_{R(i-1)}$ and $\overline{\Psi_{Li}}
\chi_{i}\Psi_{Ri}$, where the horizontal charge of all the $M_j$'s is $a$
and that of all the $\chi_i$'s is $b$.
In turn, eqs.(\ref{horcons}) produce the recursion relations
\bml
\begin{eqnarray}
 & R_{i+1} = R_{i}+(a-b) \quad \rm{for\ all} \ i ~, &  \\
 & L_{i+1} = L_{i}+(a-b) \quad \rm{for}\ i\neq 0 ~, &
\end{eqnarray}
\eml
as a simple yet quite restrictive consequence.
Such a $U(1)_{H}$ symmetry explains the stairway texture, and also
provides a rationale for all the $\chi_{i}$'s to be of the
same order of magnitude (and similarly all the
$M_{j}$'s).
However, it has the usual drawback of leaving the relative Yukawa
coupling constants arbitrary.

This motivates for $U(1)_{H}$ to be embedded into a higher
non-Abelian symmetry group.
But the required embedding is
obvious and natural: \textit{an $SU(2)_{H}$
symmetry can support, at no extra cost, the equally spaced horizontal
charges}.
The $(p+1)$ right-handed see-saw fermions $\Psi_{Ri}$ furnish the
$\underline{(p+1)}$--representation, the $p$ left-handed see-saw
fermions $\Psi_{Li}$ are assigned to the $\underline{p}$, whereas
the single standard fermion $\psi_{L0}$ is a singlet.
The $a$ and $b$ charges are then $+1/2$ and $-1/2$, respectively,
thereby suggesting that $(M,\chi)$ forms an $SU(2)_{H}$ doublet.

The real issue, however, is how to construct a mass matrix which
produces one unsuppressed, one singly-suppressed, and one
doubly-suppressed eigenvalue
\bml
\begin{eqnarray}
 m_{1} & \sim & \ell ~, \\
 m_{2} & \sim & \ell\epsilon ~, \\
 m_{3} & \sim &  \ell\epsilon^{2} ~.
\end{eqnarray}
\eml
Unfortunately, this cannot be pursued by a straightforward
concatenation of the $p=0,1,2$ sectors supported by $SU(2)_{H}$.
Recalling that the usual universal see-saw model is Left-Right
symmetric, while the generalized $p$--fold suppressing case is
not, our analysis seems to (wrongly) suggest that one should
abandon Left-Right symmetry, and instead use only $SU(2)_{H}$.
It turns out, however, that the geometric mass hierarchy naturally
hints at the larger symmetry group $SU(3)_{H}$.
The point of this paper is that \textit{the universal see-saw
geometric-like hierarchy, Left-Right symmetry, and the full
$SU(3)_{H}$ can be reconciled with each other}.

\bigskip\noindent
{\textbf{Hint 3: Geometric Mass Hierarchy and $SU(3)_{H}$}}

\medskip
Consider a geometric-like mass hierarchy of the generic type
\begin{equation}
    \ell~,~\ell\epsilon~,~\ell\epsilon^{2}~,~
 \ldots~,~\ell\epsilon^{N-1}
    \label{hierarchy}
\end{equation}
for an as yet unspecified number $N$ of standard families.
This sequence of eigenmasses can be produced from the corresponding
sequence of $p$--fold suppressing universal see-saw matrices
constructed above. But the set of fermions one needs forms an
interesting pattern.
The right-handed see-saw fermions form the set
\begin{equation}
 \Psi_{R} \sim
 \underline{1}\oplus\underline{2}
 \oplus\ldots\oplus\underline{N}
\end{equation}
of $SU(2)_{H}$ representations.
This immediately suggests, given the fact that the $\Psi_{R}$'s can be
conveniently grouped (for arbitrary $N$) within a \textit{triangular}
representation, the embedding of $SU(2)_{H}$ within $SU(3)_{H}$.
In fact, both sets of see-saw fermions form triangles in the
$(T_{3},Y)_{H}$ plane, with the $\Psi_{L}$ triangle being one
rung smaller than the $\Psi_{R}$ triangle.
We therefore find a direct connection between the observed
(approximate) geometric quark mass hierarchy and the appealing
horizontal symmetry group $SU(3)_H$.
\textit{It is the geometric-like Fermi mass hierarchy, and not
directly the total number of families, that singles out $SU(3)_{H}$}.
Ironically, the standard left-handed fermions, while being $N$ in
number, only form a collection of $SU(3)_{H}$ singlets.

To take stock of the situation: We have an $SU(2)_H$ framework that can
produce any $p$--fold suppression. However, it is not left-right
symmetric and there is no constraint on the number of families. It can
incorporate the observed geometric-like
pattern of fermion masses, but it does not \textit{predict} this pattern
uniquely. However, if
the eigenmass sequence is taken as geometric, then the pattern of $SU(2)_H$
fermion representations strongly suggests an underlying $SU(3)_H$.
The number of standard left-handed fermions is, however, still
not constrained by the `horizontal' symmetry. The
pieces of the puzzle \textit{almost} fit.

The theory we describe in the next section dovetails the geometric
universal see-saw mass hierarchy, horizontal $SU(3)_{H}$ and Left-Right
symmetry.
The resulting fermion mass matrix has an intricate texture in which
a Left-Right symmetric geometric sequence of stairway patterns can
be discerned, but not in the obvious, naive way.
The key to resolving the apparent conflict between Left-Right symmetry
and the geometric sequence of $p$--fold universal see-saw suppression
is the following set of five building blocks
\begin{equation}
 \left(
 \begin{array}{c|cc|ccc|cc|c}
    &   &    &   &    &    &    &    &\ell \\
  \hline
    &   &    &   &    &    & 0  &\ell&     \\
    &   &    &   &    &    &\chi& M  &     \\
  \hline
    &   &    & 0 &  0 &\ell&    &    &     \\
    &   &    & 0 &\chi& M  &    &    &     \\
    &   &    & r & M  &\cdot&   &    &     \\
  \hline
    & 0 &\chi&   &    &    &    &    &     \\
    & r & M  &   &    &    &    &    &     \\
  \hline
  r &   &    &   &    &    &    &    &
 \end{array}
 \right) ~,
\end{equation}
generalizing the three building block toy set eq.(\ref{toy}).
Note that the ``$\cdot$'' entry in the middle block can be as
large as $M$ without spoiling the lightest eigenvalue pattern.
This sequence is manifestly Left-Right symmetric and geometric-like.
It extends the pattern of the toy model eq.(\ref{toy}), but inserts
a novel feature $\chi$ whose significance will be explained below.
The full mass matrix thus has dimensionality $1 + 2 + 3 + 2 + 1 = 9$.

Which features of the three hints are to be retained and which
abandoned? The answer to this question is not \textit{a priori}
obvious. We now present a scheme that works.

\section{Left-Right Symmetric $SU(3)_{H}$ Theory
of the Geometric Mass Hierarchy}

\noindent
% Symmetry Group and Multiplet Assignments:
{\textbf{Symmetry Group and Multiplet Assignments:}}

\medskip
The largest representation suggested by hint 3, namely $\underline
{\frac{1}{2}{\scriptstyle N(N+1)}}$, is maintained for the
right-handed see-saw fermions $\Psi_{R}$.
In keeping with our demand of Left-Right symmetry, the left-handed
see-saw fermions $\Psi_{L}$ can now be either the vector-like choice
$\underline{{1\over2} {\scriptstyle N(N+1)}}$ or the flavor-chiral choice
$\underline{{1\over2}{\scriptstyle N(N+1)^{*}}}$.
This is of course a departure from hint 3, which would require
a $\underline{{1\over2}{\scriptstyle N(N-1)}}$ or $\underline{{1\over2}
{\scriptstyle N(N-1)^{*}}}$.
For reasons to be specified soon, we choose the flavor-chiral
option rather than the vector-like option.
Next, consider the $N$ left-handed and the $N$ right-handed standard
fermions, $\psi_{L}$ and $\psi_{R}$ respectively.
Each must constitute some, preferably irreducible, representation.
We adopt the simplest non-trivial choice of either the fundamental
representation or its conjugate (again departing from hint 3).
Since we already know from the geometric-like mass hierarchy that
the horizontal group is $SU(3)_{H}$, this forces $N = 3$.
In this way, \textit{the total number of standard families gets
correlated with the geometric-like Fermi mass hierarchy}.
We resolve the conjugation ambiguity by choosing the ``same triality''
assignments
\bml
\begin{eqnarray}
 \psi_{L}\oplus\Psi_{L} & \sim &
 \underline{3}\oplus\underline{6}^{*} ~,\\
 \psi_{R}\oplus\Psi_{R} & \sim &
 \underline{3}^{*}\oplus\underline{6} ~.
\end{eqnarray}
\eml
The combined requirements of flavor-chirality and same triality
ensure that a group theoretically attractive Higgs sector can
couple to fermions through Yukawa interactions (see later).
The flavor-chiral option is also favored on electro/nuclear
unification grounds.

We now specify the multiplet structure of our theory.
Using $\left(SU(3)_{H}|SU(3)_{C}\otimes SU(2)_{L} \otimes SU(2)_{R}
\right)_{B-L}$ notation, our scheme is spanned by the Fermi sector
\begin{equation}
    \begin{array}{c}
        q_{L}(\,\underline{3}\,|
        \,\underline{3}\,,\,\underline{2}\,,
        \,\underline{1}\,)_{+{1\over 3}}+
        Q_{L}^{u,d}(\,\underline{6}^{*}\,|
        \,\underline{3}\,,\,\underline{1}\,,
        \,\underline{1}\,)_{+{4\over 3},-{2\over 3}} \\
        q_{R}(\,\underline{3}^{*}\,|
        \,\underline{3}\,,\,\underline{1}\,,
        \,\underline{2}\,)_{+{1\over 3}}+
        Q_{R}^{u,d}(\,\underline{6}\,|
        \,\underline{3}\,,\,\underline{1}\,,
        \,\underline{1}\,)_{+{4\over 3},-{2\over 3}}
    \end{array}
    \label{fermion}
\end{equation}
accompanied by leptons, and the Higgs sector
\begin{equation}
    \begin{array}{c}
        \ell_{u,d}(\,\underline{3}\,|
        \,\underline{1}\,,\,\underline{2}\,,\,
        \underline{1}\,)_{+1,-1}+
        r_{u,d}(\,\underline{3}\,|
        \,\underline{1}\,,\,\underline{1}\,,\,
        \underline{2}\,)_{-1,+1}  \\
        \Phi_{u,d}(\,\underline{6}^{*}\,|
        \,\underline{1}\,,\,\underline{1}\,,\,
        \underline{1}\,)_{0,0}.
    \end{array}
    \label{higgs}
\end{equation}

\noindent Several remarks are in order:

\medskip
\noindent 1. All left-handed fermions (and anti-fermions) and Higgs
scalars transform horizontally \textit{alike}, namely  via
$\underline{3}\oplus\underline{6}^{*}$.
The Yukawa interactions are given by
\begin{eqnarray}
    && {\cal L}_{Y} = \left[
    \lambda_{u} (\overline{q}_{L} Q^{u}_{R} \ell^{*}_{u}+
                 \overline{q}_{R} Q^{u}_{L} r_{u})+
    \Lambda_{u} \overline{Q}^{u}_{R} Q^{u}_{L} \Phi_{u}\right]+
    \nonumber\\
    && + \left[
    \lambda_{d} (\overline{q}_{L} Q^{d}_{R} \ell^{*}_{d}+
                 \overline{q}_{R} Q^{d}_{L} r_{d})+
    \Lambda_{d} \overline{Q}^{d}_{R} Q^{d}_{L} \Phi_{d}\right]+h.c ~,
    \label{yukawa}
\end{eqnarray}
where the charge $+2/3$ sector terms are grouped between the
first pair of square brackets, and the charge $-1/3$ sector between
the second pair. The $SU(3)_H$ fermion assignments selected allow, through
$SU(3)$ group theory, the parallel structure for the Higgs bosons.

\medskip
\noindent 2. The marriage of the universal see-saw mechanism,
Left-Right symmetry, and horizontal $SU(3)_{H}$ is schematically
expressed, in the Yukawa sector, by correlating
\bml
\begin{eqnarray}
  &\begin{array}{cc}
        & \begin{array}{cc}
           (\underline{1},\underline{2}) & \,
              (\underline{1},\underline{1})
          \end{array} \\
          \begin{array}{c}
              (\underline{2},\underline{1}) \\
              (\underline{1},\underline{1})
          \end{array}
        & \left(
          \begin{array}{c|c}
              - & \,(\underline{2},\underline{1})\, \\
     \hline
              \,(\underline{1},\underline{2})\, &
              \,(\underline{1},\underline{1})\,
          \end{array}
          \right)
      \end{array}  &
\end{eqnarray}
with
\begin{eqnarray}
  &\begin{array}{cc}
        & \begin{array}{cc}
              \underline{3}\,\,\,\,\, & \underline{6}^{*}\,\,\,\,
          \end{array} \\
          \begin{array}{c}
              \underline{3} \\
              \,\,\underline{6}^{*}
          \end{array}
        & \left(
          \begin{array}{c|c}
              - & \,\underline{3}\,  \\
     \hline
              \,\,\underline{3}\,\, & \,\,\,
              \underline{6}^{*}\,
           \end{array}
        \right) ~,
    \end{array} &
\end{eqnarray}
\eml
owing its simplicity to the fact that $\underline{3}\times
\underline{6}^{*}\times\underline{6}^{*}\not\supset\underline{1}$.
While being a \textit{necessary} condition for the universal
see-saw mechanism, the absence of the $(\underline{3}\oplus
\underline{6}^{*}\,|\,\underline{2},\underline{2})$ entry from
the standard electroweak corner is not a \textit{sufficient}
condition for establishing the correct mass hierarchy.
The latter calls for a certain detailed structure at the
$(\underline{6}^{*}|\underline{1},\underline{1})$ corner.
\textit{It must be singular enough to support exactly one
unsuppressed eigenmass, and patterned in a
particular manner in order to produce a full geometric-like
hierarchy}.

\medskip
\noindent 3. The possibility of a see-saw sector analogue of $SU(2)_{L}
\otimes SU(2)_{R}$ is strongly suggested by the assignments in
eq.(\ref{fermion}).
All fermions and scalars in our scheme fall naturally into
$SU(2)^{\prime}_{L}\otimes SU(2)^{\prime}_{R}$ representations:
$Q_{L}^{u,d}$ and $r_{u,d}$ can form $(\underline{2}^{\prime},
\underline{1}^{\prime})$ multiplets, $Q_{R}^{u,d}$ and $\ell_
{u,d}$ can form $(\underline{1}^{\prime},\underline{2}^{\prime})$
multiplets, while $\Phi_{u,d}$ can live in $(\underline{2}^{\prime},
\underline{2}^{\prime})$.
In this way, the see-saw sector weak-isospin
gauge group can elegantly explain
why the electric charges of see-saw matter match those of
standard matter.
Furthermore, it motivates an electro/nuclear $[SO(10)\otimes
SO(10)^{\prime}]\otimes SU(3)_{H}$ unification scheme.

\medskip
\noindent 4. The doubling of scalars into an up-type and a down-type,
dictated by the horizontal flavor chirality for the Higgs triplets,
is recognized as a vital ingredient in any supersymmetric \cite{susy}
extension of the model. Note that the doubling of Higgs sextets $\Phi$ is
dictated by gauge symmetry only when the
$SU(2)^{\prime}_L \otimes SU(2)^{\prime}_R$
extension is invoked.

\medskip
\noindent 5. The flavor-chiral nature of $SU(3)_{H}$ suggests that the
full theory contains an $SU(3)_{H}$ anomaly,
in which case the horizontal symmetry must be global rather than local.
The spontaneous breakdown of global $SU(3)_{H}$ has interesting
consequences in regard to familon\cite{fam} phenomenology and
large-scale structure formation in cosmology\cite{texture}.

\bigskip\noindent
% Mass Matrix and VEV Pattern:
{\textbf{Mass Matrix and VEV Pattern:}}

\medskip
Having discussed the important features of the multiplet assignments,
we note that, while elegant, they do not of themselves lead to the
desired geometric-like hierarchical structure. A particular VEV pattern
is also necessary.
To fully appreciate this, we must first examine the general
$9\times 9$ mass matrix that arises from the Yukawa Lagrangian
eq.(\ref{yukawa}).
Since the mass matrix is of the same form in both of the quark
sectors, we can momentarily consider the generic mass matrix ${\cal M}$:
\begin{equation}
    \left(
    \begin{array}{ccc|cccccc}
          0 & 0 & 0 & \ell^{1} & 0 & 0 &
          0 & \frac{\ell^{3}}{\sqrt{2}}& \frac{\ell^{2}}{\sqrt{2}}\\
          0 & 0 & 0 & 0 & \ell^{2} & 0 &
          \frac{\ell^{3}}{\sqrt{2}}& 0 &\frac{\ell^{1}}{\sqrt{2}}\\
          0 & 0 & 0 & 0 & 0 & \ell^{3} &
    \frac{\ell^{2}}{\sqrt{2}}&\frac{\ell^{1}}{\sqrt{2}}& 0  \\
          \hline
          r^{1} & 0 & 0 & 0 & \Phi_{33} &
    \Phi_{22} & \Phi_{23} & 0 & 0  \\
          0 & r^{2} & 0 & \Phi_{33} & 0
    & \Phi_{11} & 0 & \Phi_{13} & 0  \\
          0 & 0 & r^{3} & \Phi_{22} & \Phi_{11} & 0 & 0 & 0 &
          \Phi_{12}  \\
          0 & \frac{r^{3}}{\sqrt{2}}& \frac{r^{2}}{\sqrt{2}} &
          \Phi_{23} & 0 & 0 &
          \Phi_{11}&-\frac{\Phi_{12}}{\sqrt{2}}&-\frac{\Phi_{13}}{\sqrt{2}}\\
          \frac{r^{3}}{\sqrt{2}}& 0 & \frac{r^{1}}{\sqrt{2}} &
          0 & \Phi_{13} & 0 &
          -\frac{\Phi_{12}}{\sqrt{2}} & \Phi_{22} &
          -\frac{\Phi_{23}}{\sqrt{2}} \\
          \frac{r^{2}}{\sqrt{2}} & \frac{r^{1}}{\sqrt{2}} & 0 &
          0 & 0 & \Phi_{12} &
          -\frac{\Phi_{13}}{\sqrt{2}}& -\frac{\Phi_{23}}{\sqrt{2}} &
          \Phi_{33}
   \end{array}
   \right) ~.
\end{equation}
The various entries denote vacuum expectation values multiplied by
Yukawa coupling constants.
These entries are labeled by the corresponding Higgs multiplet
components.
Thus,
\bml
\begin{eqnarray}
 (\ell^{1},\ell^{2},\ell^{3}) & = & \lambda \langle\ell\rangle ~, \\
 (r^{1},r^{2},r^{3}) & = & \lambda \langle r\rangle ~, \\
 (\Phi_{ij}) & = & \Lambda\langle\Phi\rangle ~.
\end{eqnarray}
\eml
The first three rows are multiplied by the three horizontal components of
$\overline{q}_L$ ($q = u,d$) in the order 1, 2, 3; the second three rows by
the
diagonal (11), (22) and (33) components of $\overline{Q}^q_L$; and the last
three rows by the off-diagonal (23), (13) and (12) components of
$\overline{Q}^q_L$, where the sextet is denoted by a $3 \times 3$ symmetric
matrix. The columns are multiplied by the components of $q_R$ and $Q^q_R$
in the corresponding order. Notice the Clebsch-Gordan coefficients
$\pm 1, \pm\frac{1}{\sqrt{2}}$; they play an important role in the model.

We take the left-handed and right-handed horizontal triplet VEVs to
be aligned.
This means that we can use a horizontal rotation to transform any
two components to zero.
We choose
\begin{eqnarray}
 \ell^{1,3} & = & 0 ~,\\
 r^{1,3} & = & 0 ~,
\end{eqnarray}
and denote
\begin{eqnarray}
 \ell^{2} & \equiv & \ell ~,\\
 r^{2} & \equiv & r ~.
\end{eqnarray}
Notice that the triplet VEVs exhibit a residual $SU(2)_{H}$
symmetry in the $(1,3)$-plane.
An $SU(2)_{H}$ transformation $\Phi\to U\Phi U^{T}$ allows us
then to set $\Phi_{13} = \Phi_{31}=0$, so that
\begin{equation}
 \Phi_{ij} \equiv
 \left(
 \begin{array}{ccc}
  z & v & 0  \\
  v & x & y  \\
  0 & y & t
 \end{array}
 \right) ~.
 \label{phi13}
\end{equation}
The generic mass matrix ${\cal M}$ becomes
\begin{equation}
    {\cal M}=\left(
    \begin{array}{ccccc|cccc}
        0 & 0 & 0 & 0 & 0 & 0 & 0 & 0 & \frac{1}{\sqrt{2}}\ell \\
        0 & 0 & 0 & 0 & \ell & 0 & 0 & 0 & 0  \\
        0 & 0 & 0 & 0 & 0 & 0 & \frac{1}{\sqrt{2}}\ell & 0 & 0 \\
        0 & 0 & 0 & 0 & -t & -x & -y & 0 & 0  \\
        0 & r & 0 & -t & 0 & -z & 0 & 0 & 0  \\
  \hline
        0 & 0 & 0 & -x & -z & 0 & 0 & 0 & -v  \\
        0 & 0 & \frac{1}{\sqrt{2}}r & -y & 0 & 0 & z &
  \frac{1}{\sqrt{2}}v & 0 \\
        0 & 0 & 0 & 0 & 0 & 0 & \frac{1}{\sqrt{2}}v & x &
  \frac{1}{\sqrt{2}}y  \\
        \frac{1}{\sqrt{2}}r & 0 &  0 & 0 & 0 & -v & 0 &
  \frac{1}{\sqrt{2}}y & t
   \end{array}
   \right).
   \label{generic9}
\end{equation}

The sextet VEV pattern is dominated by a heavy mass scale $M$.
But which entries of $\Phi_{ij}$ actually carry this scale?
Clearly, they cannot all share such a property, as otherwise all
light eigenmasses get suppressed.
Indeed, the requirement of having \textit{one and only one family
with eigenmass} $\sim \ell$, a vital mass hierarchy ingredient,
severely constrains the tenable VEV pattern.
To see the point, notice that for any given $\ell$ in the
generic mass matrix (\ref{generic9}) to be suppressed at low
energies, at least one entry along its column must be $\sim M$.
Up to a residual $SU(2)_{H}$ rotation, this leaves us with only two
options to consider:

\noindent (i) Either $v\sim M$ and $t,y,z\ll M$, leading to
\begin{equation}
 \Phi_{ij}=
 \left(
 \begin{array}{ccc}
  0 & M & 0  \\
  M & x & 0  \\
  0 & 0 & 0
 \end{array}
 \right) +\ldots ~,
\end{equation}

\noindent (ii) or $z\sim M$ and $v,y,t\ll M$, for which
\begin{equation}
 \Phi_{ij} =
 \left(
 \begin{array}{ccc}
  M & 0 & 0  \\
  0 & x & 0  \\
  0 & 0 & 0
 \end{array}
 \right) + \ldots ~.
\end{equation}
Although the value of $x$ is not directly restricted, it cannot
nonetheless to be $\sim M$, as otherwise another ${\cal O}(\ell)$
eigenmass $\displaystyle{\sim \frac{x\ell}{\sqrt{M^{2}+x^{2}}}}$
would make its appearance; this refers to both options.
Also, the second option can be rejected on phenomenological grounds.
It lacks the stairway structure eq.(\ref{bblock}), and thus is not
capable of producing a doubly suppressed light eigenmass.
One thus finally deduces that the sextet VEV pattern must be
governed by
\begin{equation}
 \Phi_{ij} =
 \left(
 \begin{array}{ccc}
  0 & M & 0  \\
  M & 0 & 0  \\
  0 & 0 & 0
 \end{array}
 \right) +\ldots ~.
\end{equation}

We now look again at eq.(\ref{generic9}) where the dominant
entry $v$ will now be relabeled as $M$, and focus on the
effects of the sub-dominant $x,y,z,t$-entries.
The geometrically hierarchical stairway texture is evident.
To be more specific, here are the see-saw sub-matrices that are
primarily responsible for the geometric mass hierarchy:

\noindent 1. The $\ell$ entry in row-2 (and similarly the $r$ entry
in column-2),
\begin{equation}
 {\cal M}_{25} =  \ell\
 ({\cal M}_{52} = r) ~,
\end{equation}
does not meet any heavy $M$-entry along its column (row)
and thus stay unsuppressed.

\noindent 2. The $\frac{1}{\sqrt{2}}\ell$ entry in row-1, and
similarly the $\frac{1}{\sqrt{2}}r$ entry in column-1, are involved
in the $2\times 2$ see-saw sub-matrices
\bml
\begin{eqnarray}
  \left(
 \begin{array}{cc}
  {\cal M}_{14} & {\cal M}_{19}  \\
  {\cal M}_{64} & {\cal M}_{69}
 \end{array}
 \right)& = & \left(
 \begin{array}{cc}
  0 & \frac{1}{\sqrt{2}}\ell  \\
  -x & -M
 \end{array}
 \right) ~,  \\
 \left(
 \begin{array}{cc}
  {\cal M}_{41} & {\cal M}_{46}  \\
  {\cal M}_{91} & {\cal M}_{96}
 \end{array}
 \right)& = & \left(
 \begin{array}{cc}
  0 & -x \\
  \frac{1}{\sqrt{2}}r & -M
 \end{array}
 \right) ~.
\end{eqnarray}
\eml
These give rise to singly suppressed eigenmasses $\frac{x\ell}
{\sqrt{2}M}$ and $\frac{xr}{\sqrt{2}M}$, respectively.

\noindent 3. The $\frac{1}{\sqrt{2}}\ell$ entry of row-3 together with
the $\frac{1}{\sqrt{2}}r$ entry of row-7 participate in the $3\times 3$
Left-Right symmetric see-saw sub-matrix
\begin{equation}
 \left(
 \begin{array}{ccc}
  {\cal M}_{33} & {\cal M}_{38} & {\cal M}_{37}  \\
  {\cal M}_{83} & {\cal M}_{88} & {\cal M}_{87}  \\
  {\cal M}_{73} & {\cal M}_{78} & {\cal M}_{77}
 \end{array}
 \right)=\left(
 \begin{array}{ccc}
  0 & 0 & \frac{1}{\sqrt{2}}\ell  \\
  0 & x & \frac{1}{\sqrt{2}}M  \\
  \frac{1}{\sqrt{2}}r & \frac{1}{\sqrt{2}}M & z
 \end{array}
 \right) ~.
\end{equation}
The corresponding light eigenmass is $\displaystyle{\frac{xr\ell}
{M^{2}}}$.

So far, we have determined that the mass hierarchy parameters are
$\displaystyle{\frac{r}{M}}$ and $\displaystyle{\frac{x}{M}}$.
But, we have to yet gain insight into the significance of $y,z,t$.
To do so, we need to discuss the mixing angle hierarchies of the
Cabibbo-Kobayashi-Maskawa (CKM) matrix.
To proceed, we set
\begin{equation}
 M\gg x,y,z\gg t ~,
\end{equation}
and evaluate the effective $5\times 5$ mass matrix ${\cal M}_{LR}$
obtained by "integrating out" the four ${\cal O}(M)$-states.
On phenomenological grounds, we must require $t$ to be very small
(see below).
Note that the parameter $t$ is special because it appears in the
effective $5\times 5$ matrix directly.
This is easy to verify from eq.(\ref{generic9}), by noticing that
the bottom-right $4\times 4$ corner contains the four heavy states.
The $M\to\infty$ limit leads to the removal of all the rows and
columns containing the $M$ entry; in this process, the parameters
$x,y,z$ also get removed, but $t$ remains.

The $5\times 5$ effective mass matrix is given by
\begin{equation}
 {\cal M}_{LR} = \left(\begin{array}{cc|ccc}
  0 & 0 & 0 & 0 & \ell \\
  0 & 0 & 0 & -\frac{1}{\sqrt{2}}\ell\frac{x}{M} &
    -\frac{1}{\sqrt{2}}\ell\frac{z}{M} \\
 0 & 0 & -\ell\frac{rx}{M^{2}} &
  \frac{1}{\sqrt{2}}\ell\frac{xy}{M^{2}} &
  -\frac{1}{\sqrt{2}}\ell\frac{yz}{M^{2}} \\
  \hline
  0 & -\frac{1}{\sqrt{2}}r\frac{x}{M} &
  \frac{1}{\sqrt{2}}r\frac{xy}{M^{2}} &
  \cdots & t + \frac{y^{2}z}{M^{2}} \\
  r & -\frac{1}{\sqrt{2}}r\frac{z}{M} &
  -\frac{1}{\sqrt{2}}r\frac{yz}{M^{2}} &
  t + \frac{y^{2}z}{M^{2}} & \cdots \\
  \end{array} \right)
  \label{MLR}
\end{equation}
where the ``$\cdots$'' entries are zero to ${\cal O}(1/M^{2})$.
The heavy fermions $Q^{33}, Q^{12}, Q^{23}, Q^{31}$ have been
decoupled.
Thus, up to ${\cal O}(1/M)$ corrections, the fermionic basis
of ${\cal M}_{LR}$ consists of $\left(q_{2}, q_{1}, q_{3},
Q^{11}, Q^{22}\right)$ [note that $q_1$ and $q_2$ have
for convenience been interchanged with respect to eq.(\ref{generic9})].

A number of observations need to be made:

\noindent 1. The effective mass matrix ${\cal M}_{LR}$ displays
the remnants of Left-Right symmetry (broken only because $\ell \neq r$).

\noindent 2. The eigenvalues are given by the magnitudes of the
anti-diagonal entries [provided that (3) below holds],
\begin{equation}
 \ell ~,~ \frac{\ell}{\sqrt{2}}\frac{x}{M} ~,~ \ell r
 \frac{x}{M^{2}} ~,~ \frac{r}{\sqrt{2}}\frac{x}{M} ~,~ r ~.
\end{equation}

\noindent 3. We need the potentially large term
\begin{equation}
 t + \frac{y^{2}z}{M^{2}} \simeq 0 ~.
 \label{t}
\end{equation}
If this term is too large, the would-be ${\cal O}(\ell)$ eigenvalue
gets see-saw suppressed.

\noindent 4. With eq.(\ref{t}) in place, the effective $3\times 3$
mass matrix ${\cal M}_{eff}$ pertaining to the $r \gg \ell$ limit
is the upper-right $3\times 3$ block
\begin{equation}
 {\cal M}_{eff}=\left(
 \begin{array}{ccc}
  0 & 0 & \ell  \\
  0 & -\frac{1}{\sqrt{2}}\ell\frac{x}{M} &
        -\frac{1}{\sqrt{2}}\ell\frac{z}{M}  \\
  -\ell\frac{rx}{M^{2}} &
   \frac{1}{\sqrt{2}}\ell\frac{xy}{M^{2}} &
   -\frac{1}{\sqrt{2}}\ell\frac{yz}{M^{2}}
 \end{array}
 \right) ~.
\end{equation}

\noindent 5. Note the unusual fermion basis of ${\cal M}_{eff}$.
The right-handed partners of the mostly `standard' left-handed
quarks $(q_{2L},q_{1L},q_{3L})$ are $(Q^{22}_{R},Q^{11}_{R},
q_{3R})$, respectively, two of which are mostly `see-saw' fermions.
The lightest family is therefore predicted to have different
$SU(2)_{R}$ gauge couplings compared to the heavier families.

\noindent 6. The light family is also special in another respect in our
scheme, because it is its own $\ell \leftrightarrow r$ partner.

\noindent 7. The left-sector mixing angles are governed by the
ratios $\displaystyle{\frac{z}{M}}$ and $\displaystyle{\frac{y}{M}}$.
To improve predictivity, we would like the
mixing parameters to be related to the mass hierarchy
parameters. To this end we introduce the requirement that
\begin{equation}
    |x| \simeq |z| ~.
    \label{z}
\end{equation}

So, we have identified the requirements eqs.(\ref{t},\ref{z})
as phenomenological necessities within the general sextet VEV
pattern.
Is there something mathematically significant about these
phenomenological constraints? Quite remarkably, there is a
simple geometric underpinning.
The required VEV pattern for each quark sector can be produced by
\bml
\begin{eqnarray}
    <\underline{3}> & = &
    (\ell\ {\rm or}\ r)
    \left(
    \begin{array}{c}
        0  \\ 1  \\ 0
    \end{array}
    \right) ~, \\
    <\underline{6}^{*}> & = &
    U\left(
    \begin{array}{ccc}
        0 & M & 0  \\
        M & 0 & 0  \\
        0 & 0 & 0
    \end{array}
    \right) U^{T} ~,
\end{eqnarray}
\eml
where $U$ is some $SU(3)_{H}$ matrix.
In other words, the realistic sextet patterns are just $SU(3)_{H}$
transformations of the dominant pattern.
The charge $2/3$ and $-1/3$ sectors differ from each other by their
relative scales $\ell_{u,d}\ll r_{u,d}\ll M$ and by the $SU(3)_{H}$
rotations $U_{u,d}$.
The departure of the matrices $U_{u,d}$ from the identity is crucial
for generating the mass hierarchy (through the $x$-parameter), and
for producing the CKM mixings (through $y$ and $z$).

The requirement $\Phi_{13} = \Phi_{31} = 0$, see eq.(\ref{phi13}), is
translated into
\begin{equation}
    U_{11}U_{32}+U_{12}U_{31}=0 ~.
\end{equation}
The most general $SU(3)_{H}$ matrix (up to diagonal phase matrices
from its left and/or right) that satisfies this takes the form
\begin{equation}
    U=\left(
    \begin{array}{ccc}
        \cos\frac{\tilde{h}}{2}\cos\alpha~ &
                -\sin{\frac{h}{2}}\cos\alpha~ & ~\cdot  \\
        \sin\frac{\tilde{h}}{2} & \cos{\frac{h}{2}} & ~\cdot  \\
        \cos\frac{\tilde{h}}{2}\sin\alpha &
        \sin{\frac{h}{2}}\sin\alpha & ~\cdot
    \end{array}
    \right) ~,
    \label{U}
\end{equation}
subject to the unitarity constraint
\begin{equation}
        \cos{2\alpha}=\frac{\displaystyle{\tan\frac{\tilde{h}}{2}}}
        {\displaystyle{\tan{\frac{h}{2}}}} ~.
\end{equation}
The ``$\cdot$'' entries, calculable by means of unitarity, are
of no relevance to our discussion.
The quantities $h$ and $\tilde{h}$ are mass hierarchy parameters,
while the mixing parameter $\alpha$ is related to the Cabibbo angle.
Remembering that the Cabibbo angle is not too small, we note that
the unitarity constraint allows $\alpha$ to stay finite in the
limit $\tilde{h}\leq h\rightarrow 0$.
To the leading order in $h$ and $\tilde{h}$, and for arbitrary
$\alpha$,
\begin{equation}
    <\Phi>\simeq
        M\left(
    \begin{array}{ccc}
        -h \cos^{2}\alpha& \cos\alpha & 0  \\
        \cos\alpha & h \cos 2\alpha& \sin\alpha  \\
         0 & \sin\alpha & h \sin^{2}\alpha
    \end{array}
    \right) ~.
        \label{6VEV}
\end{equation}
This pattern is characterized by
\begin{equation}
        t+\frac{y^{2}z}{M^{2}\cos^{2}\alpha}= {\cal O}(h^{3}) ~,
\end{equation}
which is very small (note that $\displaystyle{M\cos \alpha}$ now plays the
role of $M$).
Furthermore, the pattern is traceless,
\begin{equation}
        z + x + t = 0 ~,
\end{equation}
and if $\alpha$ itself is first-order, we observe that
\begin{equation}
        z+x \simeq {\cal O}(h\alpha^{2}) ~,
\end{equation}
which is also small.
The phenomenological constraints eq.(\ref{t}) and eq.(\ref{z})
are thus satisfied in a non-trivial way.

Substituting the VEV pattern eq.(\ref{6VEV}) into
the low-energy effective mass matrix ${\cal M}_{eff}$,
we calculate the
corresponding CKM matrix $V_{CKM}$ to find that
\begin{equation}
        \begin{array}{rcl}
                V_{11} & \simeq  & \cos(\alpha_{d}-\alpha_{u}) ~,\\
                V_{22} & \simeq  & \cos(\alpha_{d}-\alpha_{u}) ~, \\
                V_{33} & \simeq  & 1 ~,\\
                V_{12} & \simeq  & \sin(\alpha_{d}-\alpha_{u}) ~, \\
                V_{21} & \simeq  & -\sin(\alpha_{d}-\alpha_{u}) ~, \\
                V_{23} &  \simeq &
                \frac{1}{\sqrt{2}}[-h_{u}\cos{2\alpha_{u}}+
                h_{d}\cos(\alpha_{d}+\alpha_{u})] ~, \\
                V_{32} &  \simeq &
                \frac{1}{\sqrt{2}}[-h_{d}\cos{2\alpha_{d}}+
                h_{u}\cos(\alpha_{d}+\alpha_{u})] ~, \\
                V_{13} &  \simeq &
                \frac{1}{\sqrt{2}}[h_{u}\sin{2\alpha_{u}}-
                h_{d}\sin(\alpha_{d}+\alpha_{u})] ~, \\
                V_{31} &  \simeq &
                \frac{1}{\sqrt{2}}[h_{d}\sin{2\alpha_{d}}-
                h_{u}\sin(\alpha_{d}+\alpha_{u})] ~.
                \end{array}
        \label{CKM}
\end{equation}
The qualitative structure of $V_{CKM}$ is correct:
$V_{12}$ is suppressed with respect to the diagonal, $V_{23}$ and
$V_{13}$ are suppressed by the mass hierarchy parameters $h_{u,d}$,
and $V_{13}$ is smaller than $V_{23}$ by a factor of the order of
the Cabibbo angle.

Before proceeding to discuss mass and mixing angle relationships in
detail, let us re-emphasize the geometric origin of the hierarchy
parameters $h_{u,d}$ and the mixing parameters $\alpha_{u,d}$.
They arise by applying $SU(3)_H$ transformations on a simple
primordial $\langle\Phi\rangle$ form.
The hierarchies are then due to these transformations
$U_{u,d}$ being only slightly different from the identity.
This suggests that the observed pattern of quark mass and mixing,
despite its complexity, may reflect a simple geometric structure
in an internal horizontal space.

\bigskip\noindent
% Mass and Mixing Angle Relations:
{\textbf{Mass and Mixing Angle Relations:}}

\medskip
The preceding subsection revealed the VEV pattern required for
our scheme to work.
It also provided a surprising geometric interpretation for
this pattern.
Two important questions, which appear to be intertwined, are now
in order: Is our theory
quantitatively as well as qualitatively successful? What is the
origin of the VEV pattern?

 From eq.(\ref{CKM}), we see that
\begin{equation}
        V_{12} \simeq - V_{21} \simeq \alpha_{d} - \alpha_{u}.
\end{equation}
Interestingly, within the framework of our geometric mass hierarchy
model, the Cabibbo angle
\begin{equation}
        \theta_{c} = \alpha_{d} - \alpha_{u} \simeq 0.22 ~,
\end{equation}
has nothing to do with the mass hierarchy.
The hierarchy and the mixing parameters may get correlated, however,
by means of the Higgs potential.

Assuming that $\alpha_{u,d}\ll 1$, the hierarchy parameters can
be easily identified
\bml
\label{h}
\begin{eqnarray}
        h_{u} & \simeq & \pm\sqrt{2} \frac{m_{c}}{m_{t}} ~,\\
        h_{d} & \simeq & \pm\sqrt{2} \frac{m_{s}}{m_{b}} ~,
\end{eqnarray}
\eml
where the sign ambiguities remain unresolved at this level.
Equation (\ref{CKM}) then produces the first indication of a mass and
mixing angle relationship,
\begin{equation}
        V_{23} \simeq - V_{32} \simeq
        \frac{1}{\sqrt{2}}(h_{d} - h_{u})\simeq
        \frac{m_{s}}{m_{b}}\pm\frac{m_{c}}{m_{t}} ~.
        \label{23}
\end{equation}
On phenomenological grounds, we need the relative sign
between the mass ratios to be positive (see below).

The light-heavy mixing entries are given by
\bml
\label{13}
\begin{eqnarray}
        V_{13} & \simeq  &
        -\frac{1}{\sqrt{2}} h_{d} (\alpha_{u} + \alpha_{d}) +
        \sqrt{2} h_{u} \alpha_{u} ~, \\
        V_{31} & \simeq &
        -\frac{1}{\sqrt{2}} h_{u} (\alpha_{u} + \alpha_{d}) +
        \sqrt{2} h_{d} \alpha_{d} ~.
\end{eqnarray}
\eml
A second phenomenologically successful relationship is then
\begin{equation}
        V_{13} + V_{31} \simeq \frac{1}{\sqrt{2}}(h_{d} -
        h_{u})(\alpha_{d} - \alpha_{u}) \simeq V_{12} V_{23},
        \label{1331}
\end{equation}
involving mixing angles only (at this level of approximation), and
independent of the sign ambiguities.

To proceed, we need more information about the hierarchy parameters
$h_{u,d}$ and the mixing parameters $\alpha_{u,d}$.
The obvious place to seek this information is from the minimization
of the Higgs potential.
Unfortunately, the full Higgs potential is very complicated, making
it difficult to extract insight.
We can obtain very interesting results, however, from a partial
analysis, which we now present.

The full Higgs potential contains all renormalizable, gauge and
$SU(3)_{H}$ invariant terms involving the Higgs multiplets
$\ell_{u,d}$, $r_{u,d}$ and $\Phi_{u,d}$.
These terms fall into two classes: those that are sensitive to
the internal structure of the Higgs multiplets, and those that
only depend on the overall scales of the Higgs multiplets.
Our partial analysis has the following restrictions:

\noindent (i) It focuses on the internal structure sensitive class.

\noindent (ii) It assumes that the horizontal triplet VEVs are fixed
in the $(0,1,0)$ configuration.

\noindent (iii) It assumes that the sextet VEV pattern is of the
form required by the geometric ansatz.

\noindent (iv) It invokes, for simplicity, constraints motivated
by a possible $SU(2)^{\prime}_{L}\otimes SU(2)^{\prime}_{R}$
see-saw weak isospin gauge group.

The Higgs potential, with the above restrictions incorporated, can
be written in the form
\begin{equation}
        V = V_{0}+V_{int}(\alpha_{u},\alpha_{d},h_{u},h_{d}) ~,
\end{equation}
where
\begin{eqnarray}
        &&  V_{int} =- \frac{1}{4} a_{1}
        {\rm Tr}(\Phi_{u}^{\dagger} \Phi_{d}){\rm Tr}(\Phi_{d}^{\dagger}
        \Phi_{u})- \nonumber \\
        && - a_{2} {\rm Tr}(\Phi_{u}^{\dagger}
        \Phi_{d} \Phi_{d}^{\dagger} \Phi_{u})+ \nonumber \\
        && + b \langle r_{u}^{i}\rangle \langle r_{d}^{i'}\rangle
        \epsilon^{ijk}\epsilon^{i'j'k'} \Phi_{jk'}\Phi_{j'k}+
        \nonumber\\
        && + c_{1} [\langle r_{d}\rangle^{\dagger} \Phi_{u}^{\dagger}
        \Phi_{u} \langle r_{d}\rangle
        +  \langle r_{u}\rangle^{\dagger} \Phi_{d}^{\dagger}
        \Phi_{d} \langle r_{u}\rangle]+ \nonumber \\
        && + c_{2} [\langle r_{u}\rangle^{\dagger} \Phi_{u}^{\dagger}
        \Phi_{u} \langle r_{u}\rangle
        +  \langle r_{d}\rangle^{\dagger} \Phi_{d}^{\dagger}
        \Phi_{d} \langle r_{d}\rangle]+ \nonumber \\
        && + (r \to \ell) ~.
\end{eqnarray}
Expanding up to fourth order in $\alpha$ and $h$, we obtain
\begin{eqnarray}
        && V_{int} \simeq (a_{1}+a_{2})[(\alpha_{u}-\alpha_{d})^{2}
        +\frac{1}{3}(\alpha_{u}-\alpha_{d})^{4}]+ \nonumber\\
        && +2[ a_{2}(1-3\alpha_{u}^{2}-\alpha_{d}^{2}+\alpha_{u}
        \alpha_{d})+a_{1}\alpha_{u}\alpha_{d}+2 p_{u}
        \alpha_{u}^{2}]h_{u}^{2}+ \nonumber\\
        && + 2[a_{2}(1-\alpha_{u}^{2}-3\alpha_{d}^{2}+\alpha_{u}
        \alpha_{d})+a_{1}\alpha_{u}\alpha_{d}+2 p_{d}
        \alpha_{d}^{2}]h_{d}^{2}- \nonumber\\
        && -4[ a_{2}(1-2\alpha_{u}^{2}-2\alpha_{d}^{2}+\alpha_{u}
        \alpha_{d})+ \nonumber\\
        && +a_{1}\alpha_{u}\alpha_{d}+
        q(\alpha_{u}^{2}+\alpha_{d}^{2})]h_{u}h_{d}
        -\frac{4}{3}a_{2}(h_{d}-h_{u})^{4},
        \label{Vint}
\end{eqnarray}
where
\bml
\label{pdefn}
\begin{eqnarray}
        q & \equiv & \frac{1}{2} b r_{u} r_{d} ~, \\
        p_{u} & \equiv & -\frac{1}{2}(c_{1}r_{d}^{2}+c_{2}r_{u}^{2}) ~, \\
        p_{d} & \equiv & -\frac{1}{2}(c_{1}r_{u}^{2}+c_{2}r_{d}^{2}) ~,
\end{eqnarray}
\eml
having inputted $|r|\gg |\ell|$.

To minimize this potential, we keep the $\alpha$'s and $h$'s small,
and identify a region of parameter space that is phenomenologically
interesting.
It turns out that the relevant region in parameter space is where the
$p$'s dominate the quartic couplings
\begin{equation}
        |p_{u,d}| \gg |q|,\ |a_{1,2}|,
\end{equation}
leading to
\begin{eqnarray}
        V_{int} & \simeq & (a_{1}+a_{2})(\alpha_{u}-\alpha_{d})^{2}
        +2a_{1} (h_{u}-h_{d})^{2}+ \nonumber \\
        & + & 4(p_{u}\alpha_{u}^{2}h_{u}^{2}+
        p_{d}\alpha_{d}^{2}h_{d}^{2}) ~.
\end{eqnarray}

Two of the four minimization equations then lead to the remarkable
relations
\begin{equation}
        \frac{h_{u}}{h_{d}} = \frac{\alpha_{u}}{\alpha_{d}}  = -
        \left(\frac{p_{d}}{p_{u}}\right)^{1/3} ~.
        \label{pupd}
\end{equation}
Given that the ratio of the $p$'s is positive (as the potential is
bounded from below), one very important consequence is that
\begin{equation}
        \frac{h_{u}}{h_{d}}<0 ~.
        \label{minus}
\end{equation}
This resolves the sign ambiguity of eq.(\ref{23}), giving rise
to the phenomenologically successful relation
\begin{equation}
        V_{23} \simeq - V_{32} \simeq \frac{m_{s}}{m_{b}} +
        \frac{m_{c}}{m_{t}} ~.
\end{equation}

Combining eq.(\ref{pupd}) with the $\theta_{c}$ expression,
we get
\bml
\begin{eqnarray}
        \alpha_{d} & = & \frac{h_{d}}{h_{d}-h_{u}}\theta_{c} ~, \\
        \alpha_{u} & = & \frac{h_{u}}{h_{d}-h_{u}}\theta_{c} ~.
\end{eqnarray}
\eml
Substituting these results into the light-heavy mixing terms,
we can now obtain individual relations for $V_{13}$
and $V_{31}$, namely
\begin{eqnarray}
        && \displaystyle{V_{13} \simeq \frac{1}{\sqrt{2}}(h_{d}+
        2 h_{u}) \theta_{c}\simeq \left( 2\frac{m_{c}}{m_{t}} -
        \frac{m_{s}}{m_{b}}\right)\theta_{c}} ~, \\
        && \displaystyle{V_{31} \simeq - \frac{1}{\sqrt{2}}(h_{u}+
        2 h_{d}) \theta_{c}\simeq \left( 2\frac{m_{s}}{m_{b}} -
        \frac{m_{c}}{m_{t}}\right)\theta_{c}} ~,
\end{eqnarray}
and establish the $\theta_{c}$--independent relation
\begin{equation}
        \frac{V_{13}}{V_{31}} \simeq
        \frac{\displaystyle{2\frac{m_{c}}{m_{t}}-\frac{m_{s}}{m_{b}}}}
        {\displaystyle{2\frac{m_{s}}{m_{b}}-\frac{m_{c}}{m_{t}}}} ~.
\end{equation}
The relative minus signs in these equations are a
non-trivial consequence of eq.(\ref{minus}).

Our final relation interrelates quark masses. From
eq.(\ref{pdefn}) we learn that
\begin{equation}
        \frac{p_{d}}{p_{u}} =
        \left(\frac{c_{1} r_{u}^{2} + c_{2} r_{d}^{2}}
        {c_{1} r_{d}^{2} + c_{2} r_{u}^{2}}\right) ~.
        \label{pr}
\end{equation}
Our model does not tell us why $|h_{u}|\ll |h_{d}|$.
It points, however, towards the parameter space region
\begin{equation}
        c_{1} \gg c_{2} .
\end{equation}
In this case,
\begin{equation}
        \frac{p_{d}}{p_{u}} \simeq
        \left(\frac{r_{u}}{r_{d}}\right)^{2} ~,
\end{equation}
leading to
\begin{equation}
        \frac{h_{u}}{h_{d}} \simeq
        -\left(\frac{r_{u}}{r_{d}}\right)^{2/3} ~.
\end{equation}
Furthermore, invoking the see-saw weak isospin restriction which implies
$\lambda_{u}=\lambda_{d}$ and $\Lambda_{u}=\Lambda_{d}$, and recalling
the eigenvalue spectrum, we arrive at
\begin{equation}
        \frac{r_{u}}{r_{d}} \simeq
        \frac{m_{u}/m_{c}}{m_{d}/m_{s}} ~.
        \label{rr}
\end{equation}
Combining eq.(\ref{rr}) and eq.(\ref{pupd}),
an unusual mass relation makes its appearance:
\begin{equation}
        \frac{m_{u}}{m_{d}} \simeq \left(\frac{m_{b}}{m_{t}}\right)^{3/2}
        \left(\frac{m_{c}}{m_{s}}\right)^{5/2} ~.
        \label{u/d}
\end{equation}
It is remarkable that the light quark mass ratio gets
fixed by the heavy quark mass ratios.

\section{Discussion and Conclusions}

The scheme developed in this paper provides, we believe, a
non-trivial framework that may contribute to a resolution of the
flavor problem.
It is successful in several very important respects:

\noindent 1. It seamlessly merges the universal see-saw mechanism,
Left-Right symmetry and horizontal symmetry in a coherent gauge
theoretic structure.

\noindent 2. It establishes a connection between the experimentally
observed (approximate) geometric mass hierarchy and the appealing
horizontal symmetry group $SU(3)_{H}$.

\noindent 3. Further, it establishes a connection between the geometric
mass hierarchy and threefold family replication.

\noindent 4. The symmetry breaking pattern is interpreted in
geometric terms.
The observed mass and mixing angle hierarchies reflect a simple
horizontal structure.

\noindent 5. The CKM matrix has the correct qualitative pattern.

\noindent 6. The model gives rise to phenomenologically successful
mass-ratio (rather than square-mass-ratio) mixing relations, and
fixes the $\displaystyle{\frac{m_{u}}{m_{d}}}$ mass ratio.

Using the heavy quark masses as input, we can probe our predictions
numerically.
Given the following $m_{q}(m_{Z})$ values\cite{qmass}
\begin{equation}
        \begin{array}{ccr}
                m_{s} & \approx & 93 ~{\rm MeV} ~, \\
                m_{c} & \approx & 677 ~{\rm MeV} ~,\\
                m_{b} & \approx & 3.0 ~{\rm GeV} ~,\\
                m_{t} & \approx & 181 ~{\rm GeV} ~,
        \end{array}
\end{equation}
and the Cabibbo angle $\theta_{c}\simeq 0.22$, we estimate the other
CKM mixings to be
\begin{eqnarray}
        && {\displaystyle V_{cb}\simeq-
        V_{ts} \simeq \frac{m_{s}}{m_{b}}+
        \frac{m_{c}}{m_{t}} \simeq 0.035} ~,\\
        && {\displaystyle V_{ub} \simeq \left(2\frac{m_{c}}{m_{t}}-
        \frac{m_{s}}{m_{b}}\right)\theta_{c}\simeq 0.005} ~,\\
        && {\displaystyle V_{td} \simeq \left(2\frac{m_{s}}{m_{b}}-
        \frac{m_{c}}{m_{t}}\right)\theta_{c}\simeq 0.013} ~.
\end{eqnarray}
These values should be compared with the experimental averages (only
the first two of which have been measured) $0.037$ and $0.004$,
respectively.

Given the $m_{q}(1~{\rm GeV})$ values\cite{qmass} of
\begin{equation}
        \begin{array}{ccr}
                m_{s} & \approx & 175 ~{\rm MeV} ~, \\
                m_{c} & \approx & 1.51 ~{\rm GeV} ~,\\
                m_{b} & \approx & 7.18 ~{\rm GeV} ~,\\
                m_{t} & \approx & 475 ~{\rm GeV} ~,
        \end{array}
\end{equation}
we predict a light quark mass ratio of
\begin{equation}
        \frac{m_{u}}{m_{d}} \simeq \left(\frac{m_{b}}{m_{t}}
        \right)^{3/2}\left(\frac{m_{c}}{m_{s}}\right)^{5/2}
        \simeq 0.44 ~.
\end{equation}
This result is to be compared\cite{u/d} with $0.55$ calculated by
Gasser-Leutwyler and Dominguez-deRafael, with $0.29$ advertized by
Donoghue-Holstein-Wyler, and the $0.4$ value derived by Narison.

There are also several important open problems facing our model.
Of immediate relevance to the discussion in this paper, it is clear
that the symmetry breaking analysis is incomplete.
The complexity of the Higgs systems needed to break attractive and
powerful higher symmetries is a general model building problem.
In our case, while the Higgs boson multiplet structure has some
appealing features, the analysis of the general Higgs potential is
bedeviled by a proliferation of parameters.
In this regard, the remarkable phenomenological success of the
partial analysis presented above is encouraging.
However, we do not as yet know whether the required VEV pattern
exists as an absolute minimum of the Higgs potential for a range
of parameters.
To rephrase the question, we can ask: What is the simplest Higgs
system that can furnish the VEV pattern we need? Is the system
discussed in this paper sufficient, or are other multiplets needed?
In particular, does supersymmetry, with its strong constraints
on the Higgs potential, have an important role to play? (The cubic
nature of superpotentials suggests an intriguing connection between
$SU(3)_{H}$ invariance and the number of Yukawa legs).
It is important to know how robust are our mass and mixing angle
relations, because the partial Higgs potential minimization procedure
was central in the derivation of some of them.

The role of even higher symmetries is also an open problem.
We have seen that see-saw weak-isospin group $SU(2)^{\prime}_
{L}\otimes SU(2)^{\prime}_{R}$ is immediately suggested by the
multiplet structure of the theory.
If it exists, it also requires that different Higgs sextets couple
to the charge $+2/3$ and $-1/3$ sectors, which we invoked for
phenomenological reasons.
There is a further suggestive extension to grand unified
$SO(10)\otimes SO(10)^{\prime}$, consistent with the flavor-chiral
nature of $SU(3)_{H}$.
If $SU(3)_{H}$ is truly a global symmetry, as suggested by its
flavor-chirality and consequent anomalies, then its deeper origin
is another open problem.
Indeed, the essential use of the $\underline{3} \oplus \underline{6}
^{*}$ representation of $SU(3)_{H}$ hints at a possible underlying
composite model, for exactly the same reasons that Gell-Mann--Ne'eman
SU(3) suggested the existence of quarks.
The simple group-theoretic result that
\begin{equation}
        \underline{3}^{*}\otimes\underline{3}^{*}=
        \underline{3}\oplus\underline{6}^{*}
\end{equation}
may imply that the multiplet structure we have used can be constructed
from two-body bound states of preons in the $\underline{3}^{*}$
representation of $SU(3)_{H}$.

The model presented in this paper does not explain the hierarchy
between the top and bottom quark masses.
Recall that there is one unsuppressed eigenvalue $\ell$ per charge
sector, which indeed matches the observed value of $m_{t}$ beautifully.
However, the relatively low mass for the bottom quark can only be
incorporated through a Yukawa coupling constant hierarchy
\begin{equation}
        \frac{\lambda_{d}}{\lambda_{u}} = \frac{m_{b}}{m_{t}} ~.
\end{equation}
The theory as it stands does not address the question of mass
splitting between weak-isospin partners.
Several possible approaches to this deep issue are apparent:
Perhaps the mass parameter $t$ in the
effective $5\times 5$ mass matrix ${\cal M}_{LR}$ is significantly
larger for the charge $-1/3$ sector than it is for the charge $+2/3$
sector.
In that case, the would-be order $\ell$ eigenvalue becomes see-saw
suppressed.
Another perspective on the top-bottom mass splitting issue is afforded
by the question: Does custodial SU(2) have a role to play?
This accidental symmetry of the minimal standard model Higgs
potential, if enforced, leads to mass degeneracy between
weak-isospin partner fermions.
Interestingly, custodial SU(2) can sometimes be identified with
$SU(2)_{R}$\cite{custodial}.

We have not explicitly discussed leptons in this paper, though the
most straightforward extension of our model to leptons is obvious.
The full neutral-lepton mass matrix is then an $18\times 18$ matrix
which should have interesting calculable features.
In the simple universal see-saw model, the Majorana see-saw mechanism
for neutrinos is automatic, leading to enhanced suppression for neutrino
masses.
A similar phenomenon is expected here, but the question of whether the
light neutrinos are necessarily hierarchical needs to be answered.
The mass splitting between quarks and leptons is another deep issue.
One well-known approach to this problem lies in the exploration of
symmetries between quarks and leptons, for example grand unification,
Pati-Salam
SU(4), and discrete quark-lepton symmetry\cite{dql}.

Finally, there is the question of experimental testability.
The phenomenology of the familon-like Goldstone bosons arising from
the breakdown of global $SU(3)_{H}$ needs to be analyzed.
Also, the production of cosmological texture in the early universe and
its role in large-scale structure formation may lead to a cosmological
test of spontaneously broken global $SU(3)_{H}$.
The mass and mixing angle relations we derived will be tested further
as more data is gathered, particularly for the top-quark CKM elements
$V_{32}$ and $V_{31}$.
If right-handed weak interactions should be discovered, then our theory
predicts that the up and down quarks will couple more strongly to $W_{R}$
and $Z_{R}$ than will charm, strange, top and bottom.
Finally, the neutral-lepton mass matrix may have testable consequences
for neutrinos, an issue we intend to explore in the future.

\acknowledgments
AD would like to thank Professor Bruce McKellar and the School of
Physics at The University of Melbourne for the enjoyable semester
during which much of this research was carried out.
RRV would like to thank the Physics Department at Ben-Gurion
University for their kind hospitality during the completion of
this work. RRV is supported by the Australian Research Council.

\end{document}